\documentclass[doublecol]{epl2}
\usepackage{graphicx}
\usepackage{bm}
\usepackage{amssymb}
\usepackage{epsfig}
\newcommand{\BE}{\begin{equation}}
\newcommand{\EE}{\end{equation}}
\title{Cosmic Background Radiation and `ether-drift' experiments}
\shorttitle{Cosmic Background Radiation and `ether-drift' experiments }

\author{M. Consoli $^{1}$, A. Pluchino $^{2,1}$, A. Rapisarda $^{2,1}$}

\institute{
  \inst{1} INFN, Sezione di Catania - Via S. Sofia 64, I-95123
Catania, Italy\\
  \inst{2} Dipartimento di Fisica e Astronomia, Universit\`a di
Catania - Via S. Sofia 64, I-95123
Catania, Italy\\
}

\pacs{98.70.Vc}{Background radiations}
\pacs{07.60.Ly}{Interferometers} \pacs{89.75.Fb} {Structures and
organization in complex systems}

\abstract{`Ether-drift' experiments have played a crucial role for
the origin of relativity. Though, a recent re-analysis shows that
those original measurements where light was still propagating in
gaseous systems, differently from the modern experiments in vacuum
and in solid dielectrics, indicate a small universal anisotropy
which is naturally interpreted in terms of a non-local thermal
gradient. We argue that this could possibly be the effect, on weakly
bound gaseous matter, of the temperature gradient due to the Earth's
motion within the Cosmic Background Radiation (CBR). Therefore, a
check with modern laser interferometers is needed to reproduce the
conditions of those early measurements with today's much greater
accuracy. We emphasize that an unambiguous confirmation of our
interpretation would have far reaching consequences. For instance,
it would also imply that all physical systems on the moving Earth
are exposed to a tiny energy flow, an effect that, in principle,
could induce forms of self-organization in matter.}

\begin{document}

\maketitle

\section{Premise}

Over the years, particular efforts have been devoted to improve the
sensitivity of those `ether-drift' experiments which look for the
possible existence of a preferred reference frame through an
anisotropy of the two-way velocity of light $\bar{c}_\gamma(\theta)$
(for a general review see e.g. \cite{applied}). This is the only one
that can be measured unambiguously and is defined in terms of the
one-way velocity $c_\gamma(\theta)$ as \BE \label{first}
\bar{c}_\gamma(\theta)={{2~c_\gamma(\theta)c_\gamma(\pi
+\theta)}\over{c_\gamma(\theta) +c_\gamma (\pi +\theta)}} \EE Here
$\theta$ represents the angle between the direction of light
propagation and the Earth's velocity with respect to a hypothetical
preferred frame $\Sigma$. By defining the anisotropy  \BE
\Delta\bar{c}_\theta =
\bar{c}_\gamma(\pi/2+\theta)-\bar{c}_\gamma(\theta)\EE the most
recent result from Nagel et al. \cite{nagel} amounts to a fractional
accuracy $(|\Delta\bar{c}_\theta| /c)\lesssim 10^{-18}$. With this
new measurement, by looking at their Fig.1 where all ether-drift
experiments are reported, one gets the impression of a steady,
substantial improvement over the original 1887 Michelson-Morley
\cite{MM} result $(|\Delta\bar{c}_\theta| /c)\lesssim 10^{-9}$.

Though, this first impression might be misleading. The various
measurements were performed in different conditions, i.e. with light
propagating in gaseous media (as in
\cite{MM,Miller,Illingworth,Joos}) or in a high vacuum (as in
\cite{brillet,newberlin,schillernew}) or inside dielectrics with a
large refractive index (as in \cite{fox,nagel}) and there could be
physical reasons which prevent such a straightforward comparison. In
this case, the difference between old experiments (in gases) and
modern experiments (in vacuum or solid dielectrics) might not depend
on the technological progress only but also on the different media
that were tested.

Another possible objection concerns the traditional analysis of the
data. The model assumed so far of slow, periodic time modulations,
associated with the Earth's rotation and its orbital revolution,
derives from simple spherical trigonometry. Here, there might be a
logical gap. The relation between the macroscopic Earth's motion and
the microscopic propagation of light in a laboratory depends on a
complicated chain of effects and, ultimately, on the physical nature
of the vacuum. By comparing with the motion of a body in a fluid,
the standard view corresponds to a form of regular, laminar flow
where global and local velocity fields coincide. However, some
arguments (for a list of references see \cite{physica}) suggest that
the vacuum might rather resemble a turbulent fluid where large-scale
and small-scale flows are only indirectly related. In this other
perspective, the macroscopic Earth's motion could just give the
order of magnitude by fixing the typical boundaries for a
microscopic velocity field which is irregular and intrinsically non
deterministic. Although it cannot be computed exactly, one could
still estimate its statistical properties by numerical simulations
\cite{plus,physica}. To this end, one could assume forms of
turbulence or intermittency which, as in most models, become
statistically isotropic at small scales. This could easily explain
the irregular character of the data because, whatever the
macroscopic Earth's motion, the average of all vectorial quantities
(such as the Fourier coefficients extracted from a fit to the
temporal sequences in modern experiments or the fringe shifts of the
old experiments) would tend to zero by increasing more and more the
statistics. In this framework, is not surprising that from an
instantaneous signal of given magnitude one ends up with smaller and
smaller averages. This trend, by itself, might not imply that there
is no physical signal.

Now, by taking into account these two ingredients, namely a) the
specificity of the various media and b) the possibility of a
genuine, but irregular, physical signal, there are substantial
changes in the interpretation of the experiments. We believe that
the main conclusions of this re-analysis, and the possible ultimate
implications, are sufficiently important to be summarized in a
concise form and thus brought to the attention of a wide audience.

\section{CBR and ether-drift experiments}

Let us first observe that the discovery of an anisotropy of the
Cosmic Background Radiation (CBR) \cite{mather,smoot} has introduced
an important new element. Indeed, the standard interpretation of its
dominant dipole component (the CBR kinematic dipole \cite{yoon}) is
in terms of a Doppler effect due to the motion of the solar system
with average velocity $v\sim 370$ km/s toward a point in the sky of
right ascension $\alpha \sim 168^o$ and declination $\delta\sim
-7^o$. This makes the existence of a preferred reference frame more
than a simple possibility.

\begin{figure}[ht] \psfig{figure=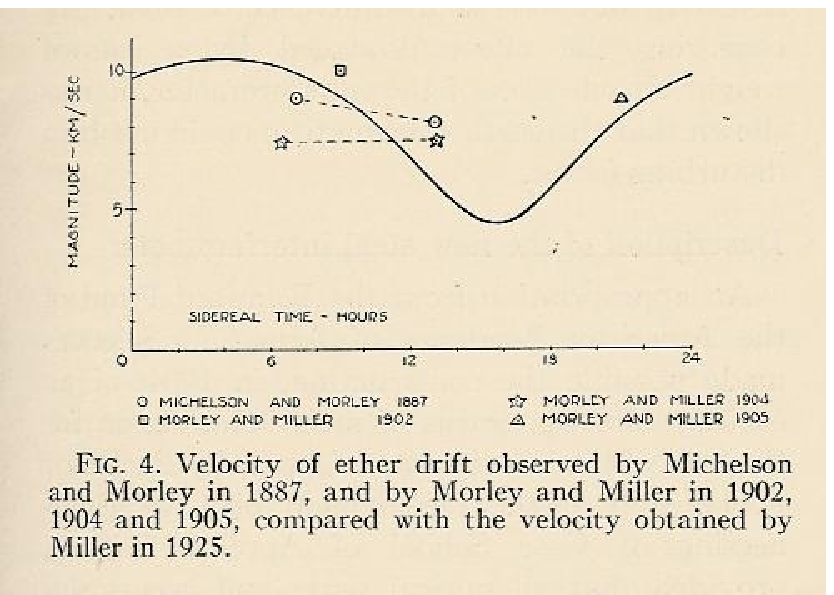,height=6 true
cm,width=8.8 true cm,angle=0} \caption{\it The velocities obtained
with Eq.(\ref{classical}) in various experiments, as reported by
Miller \cite{Miller}.}\label{fig_miller}
\end{figure}

In spite of this, it is generally assumed that this motion cannot be
detected in a laboratory by optical measurements. This belief
derives precisely from the ether-drift experiments, at least when
interpreted as a long sequence of `null results' with better and
better systematics. Still, over the years, greatest experts
\cite{Hicks,Miller} have seriously questioned the traditional null
interpretation of the early experiments. In their opinion, the small
residuals should not be neglected. To have an idea of their
magnitude, let us recall that, at the beginning, the fringe shifts
produced by the rotation of the interferometers were analyzed by
using the classical formula \BE\label{classical} (\Delta
\lambda/\lambda)_{\rm class}\sim (L/\lambda)(v^2 /c^2) \EE where $v$
is the projection of the Earth's velocity in the plane of the
interferometer, $L$ the length of the optical path and $\lambda$ the
light wavelength. Quantitatively, the very early measurements are
summarized in Fig.\ref{fig_miller} (from ref.\cite{Miller}) where
the velocities obtained with Eq.(\ref{classical}) in various
experiments are reported and compared with a smooth curve fitted by
Miller to his own results as function of the sidereal time. These
experimental velocities, lying in the range $7\div 10$ km/s, imply
that the fringe shifts in the various experimental sessions were
about $10\div 20$ times smaller than those expected classically for
the Earth's orbital value $v=30$ km/s (the minimum anticipated drift
velocity).

At the same time, although much smaller than the expected value, the
measured fringe shifts were often non negligible \cite{Miller,plus}
as compared to the extraordinary accuracy of the interferometers.
This suggests that, in some alternative framework, the small, and
irregular, effects could acquire a definite physical meaning. So
far, their interpretation has been in terms of unimportant, mainly
thermal, disturbances. However, what about a `non-local' temperature
effect? An observer moving through the CBR would see different
temperatures in different directions and, in this case, the
situation could change completely. This also suggests to concentrate
the attention on the experiments in gaseous systems because the
elementary constituents of such weakly bound matter can be set in
motion by extremely small thermal gradients.

\begin{table*}
\caption {\it The average velocity observed (or the limits placed)
by the classical ether-drift experiments in the alternative
interpretation where the fringe shifts are given by
Eq.(\ref{fringe}) and the relation between the observable $v_{\rm
obs}$ and the kinematical $v$ is governed by Eq.(\ref{vobs}). For
the dots in the Michelson-Pease-Pearson case we address the reader
to ref.\cite{plus}. }
\begin{center}
\begin{tabular}{clll}
\hline Experiment &gas in the interferometer
&~~~~$v_{\rm obs}({\rm km/s})$ & ~~~~$v$({\rm km/s})\\
\hline
Michelson-Morley(1887)   & ~~~~~~~~~~ air & ~~~~ $8.4^{+1.5}_{-1.7}$&~~~$349^{+62}_{-70}$ \\
Morley-Miller(1902-1905)  & ~ ~~~~~~~~~air& ~~~~ $8.5\pm 1.5$ & ~~ $ 353 \pm 62$ \\
Kennedy(1926)  & ~~~~~~~~~~~helium &~~~~~~$<5 $  &  ~~~~$<600 $\\
Illingworth(1927) & ~~~~~~~~~~~helium & ~~~~~$3.1 \pm 1.0$  &~~~$370 \pm 120$ \\
Miller(1925-1926) & ~~~~~~~~~~~air & ~~~~~$8.4^{+1.9}_{-2.5}$ &~~~$349^{+79}_{-104}$ \\
Michelson-Pease-Pearson(1929)& ~~~~~~~~~~~air &~~~~~$4.5\pm... $  &  ~~~$ 185 \pm ... $\\
Joos(1930)  &~~~~~~~~~~~helium&~~~~$\ 1.8^{+0.5}_{-0.6} $  & ~~ $330^{+40}_{-70} $\\
\hline
\end{tabular}
\end{center}
\end{table*}
To estimate this effect, let us recall that, due to the motion of an
observer with velocity $v$, a pure black-body spectrum of
temperature $T_o$ becomes Doppler shifted in the various directions
$\theta$ according to the relation ($\beta=v/c)$ \BE
T(\theta)={{T_o\sqrt{1-\beta^2}}\over{1- \beta \cos \theta} } \EE
Therefore, if one sets $T_o \sim $ 2.7 K and $\beta\sim 0.0012$ as
for $v=$ 370 km/s, there is an angular variation \BE
\label{CBR}\Delta T(\theta) \sim T_o \beta \cos\theta \sim \pm 0.003
~{\rm K} \EE A more accurate estimate for an ether-drift experiment
would first require to replace the value $v=$ 370 km/s with its
projection in the plane of the interferometer and then evaluate the
effects on the observation site. We have not attempted this
non-trivial task. However, for Miller's observations this analysis
was carried out by Kennedy, Shankland (see p.175 of
\cite{shankland}, in particular the footnote$^{16}$) and Joos
\cite{joos2}. Their conclusion was that periodic temperature
variations of about $\pm 0.001$ K or $\pm 0.002$ K in the air of the
optical arms could be responsible for Miller's average fringe
pattern. Now, on the one hand, these temperature values agree well
with Eq.(\ref{CBR}). On the other hand, such interpretation of the
residual effects would also fit with Miller's conclusion
\cite{miller2} that the needed temperature variations could not be
due to a uniform heating (or cooling) of the laboratory but should
have been those produced by a directional effect, as it would be
with the CBR dipole.

With this premise, it becomes important to check if the small
residuals indicate a non-local phenomenon that could be interpreted
as a universal temperature gradient. To this end, we summarize in
Table 1 the main results of ref.\cite{plus} which represents the
most complete analysis performed so far of the classical experiments
in gaseous media (Michelson-Morley \cite{MM}, Miller \cite{Miller},
Illingworth \cite{Illingworth}, Joos \cite{Joos}...). For the
typical projections $v$ associated with an Earth's velocity of 370
km/s, by introducing the gas refractive index ${\cal N}=1+
\epsilon$, the experimental fringe shifts produced by the rotation
of the interferometers were found to scale as \BE \label{fringe}
(\Delta\lambda/\lambda)_{\rm EXP} \sim (L/\lambda)(v^2_{\rm
obs}/c^2) \EE with an `observable' velocity \BE \label{vobs}
v^2_{\rm obs} \sim 2\epsilon v^2 \EE Notice that the effect vanishes
in the $\epsilon \to 0$ limit, as expected when the velocity of
light $c_\gamma$ approaches the basic parameter $c$ entering Lorentz
transformations. Thus one gets $(v^2_{\rm obs} /c^2)\lesssim
10^{-9}$ for air at atmospheric pressure, where ${\cal N}\sim
1.00029$, or $(v^2_{\rm obs} /c^2)\lesssim 10^{-10}$ for helium at
atmospheric pressure, where ${\cal N}\sim 1.000035$. To appreciate
the strong suppression effect, one should compare with the
corresponding classical prediction Eq.(\ref{classical}). For
instance, for air, the fringe shifts for $v=$ 370 km/s are about 10
times smaller than those expected classically for the much lower
velocity $v=$ 30 km/s. For gaseous helium, the effect is even 100
times smaller. We believe that the good agreement among the various
determinations of $v$ in Table 1 provides enough evidence for the
existence of a non-local effect that should be understood.

\section{Derivation of the observed anisotropy}

Within the traditional thermal interpretation, the ultimate
explanation of the observed universal anisotropy proportional to
$\epsilon (v/c)^2$ was searched for \cite{plus,foop} in the
fundamental energy flow which, on the basis of general arguments, is
expected in a quantum vacuum which is not exactly Lorentz invariant
and thus sets a preferred reference frame. However, the agreement
between Eq.(\ref{CBR}) and the old estimates of Joos, Kennedy and
Shankland introduces now a new argument and provides the most
natural interpretation in terms of the CBR itself.

To try to understand Eqs.(\ref{fringe}) and (\ref{vobs}), one can
first start from standard assumptions, namely:

~~~i) light anisotropy should vanish when both the observer and (the
container of) the medium where light propagates are taken at rest in
the hypothetical preferred frame $\Sigma$, for instance the system
where the CBR looks exactly isotropic

~~ii) light anisotropy should also vanish if light propagates in an
ideal vacuum, i.e. for a medium refractive index ${\cal N} = 1$ so
that $c_\gamma$ coincides with $c$

This means that, in the physical case where instead both the
observer and (the container of) the medium are at rest in the
laboratory $S'$ frame, any possible anisotropy should vanish
identically in the limit of velocity $v= 0$ when $S'\equiv \Sigma$.
Therefore, if we restrict our analysis to the region $ {\cal N} =1+
\epsilon$ of gaseous media, one can expand in the two small
parameters $\beta= v/c$ and $\epsilon= {\cal N} -1$. Then, any
possible anisotropy will start to ${\cal O}(\epsilon\beta)$ for the
one-way velocity $c_\gamma(\theta)$ and to ${\cal
O}(\epsilon\beta^2)$ for the two-way velocity
$\bar{c}_\gamma(\theta)$ which, by its very definition, is invariant
under the replacement $\beta \to -\beta$. At the same time, for any
fixed $\beta$, $\bar{c}_\gamma(\theta)$ is also invariant under the
replacement $\theta \to \pi +\theta$. Thus, to lowest non-trivial
level ${\cal O}(\epsilon\beta^2)$, one finds the general expression
\begin{eqnarray}
\label{legendre} \bar{c}_\gamma(\theta) \sim {{c}\over{ {\cal N} }}
\left[1- \epsilon\beta^2
\sum^\infty_{n=0}\zeta_{2n}P_{2n}(\cos\theta)
  \right]
\end{eqnarray}
Here, to account for invariance under $\theta \to \pi +\theta$, the
angular dependence has been given as an infinite expansion of
even-order Legendre polynomials with arbitrary coefficients
$\zeta_{2n}={\cal O}(1)$.

A crucial point for the thermal interpretation is that
Eq.(\ref{legendre}) admits a dynamical basis. In fact, exactly the
same form is obtained \cite{plus} (see also Appendix 1 of
\cite{foop}) if, in the $S'$ frame, there were convective currents
of the gas molecules associated with an Earth's absolute velocity
$v$. Both derivations clearly differentiate gaseous systems from
solid and liquid dielectrics (where instead ${\cal N}$ differs
substantially from unity) and, therefore, one can understand the
difference with strongly bound matter, as in the Shamir-Fox
experiment \cite{fox}. Being aware that the classical measurements
could be proportional to $\epsilon (v/c)^2$, they selected a medium
where the effect of the refractive index would have been enhanced
(i.e. perspex where ${\cal N}\sim 1.5$). Since this enhancement was
not observed, they concluded that the experimental basis of special
relativity was strengthened. However, with a thermal interpretation,
one can reconcile the different behaviors because in solid
dielectrics a small temperature gradient would mainly dissipate by
heat conduction without generating any appreciable particle motion
or light anisotropy in the rest frame of the apparatus.

Now, Eq.(\ref{legendre}) is exact to the given accuracy and predicts
the right order of magnitude $\epsilon (v/c)^2$ of the observed
anisotropy. Therefore, by leaving out the first few $\zeta's$ as
free parameters in the fits, one could directly compare with the
experimental data. Still, there is one more derivation of the
$\epsilon \to 0$ limit with a preferred frame which, on the basis of
other symmetry arguments, permits to get rid of the unknown
coefficients in (\ref{legendre}) and to deduce Eqs.(\ref{fringe})
and (\ref{vobs}). The reason is that the transformation matrix which
connects the space-time metric $g^{\mu\nu}$ for light propagation in
the laboratory $S'$ frame to the reference isotropic metric
$\gamma^{\mu\nu}={\rm diag}({\cal N}^2,-1,-1,-1)$ in the preferred
$\Sigma$ frame, is a two-valued function for ${\cal N}\to 1 $. As
shown in Appendix 2 of ref.\cite{foop}, by taking into account this
subtlety, there are two solutions: either
$g^{\mu\nu}=\gamma^{\mu\nu}$ or $g^{\mu\nu} \sim \eta^{\mu\nu} +
2\epsilon u^\mu u^\nu$ where $\eta^{\mu\nu}$ is the Minkowski tensor
and $u^\mu$ the dimensionless $S'$ 4-velocity. With the latter
choice, from the condition $p_\mu p_\nu g^{\mu\nu}=0$, by defining
$c_\gamma(\theta)$ from the ratio $p_0/|{\bf p}|$ and using
Eq.(\ref{first}), one finds a two-way velocity
\begin{eqnarray}
\label{twoway0}
       \bar{c}_\gamma(\theta)
       &\sim& (c/ {\cal N})\left[1-\epsilon\beta^2\left(2 -
       \sin^2\theta\right) \right]
\end{eqnarray}
which corresponds to setting in Eq.(\ref{legendre}) $\zeta_0=4/3$,
$\zeta_{2}= 2/3$ and all $\zeta_{2n}=0$ for $n
> 1$. Eq.(\ref{twoway0}) is a definite
realization of the general structure in (\ref{legendre}) and
provides a partial answer to the problem of calculating the
$\zeta's$ from first principles. As such, it represents a model to
compute the time difference for light propagation back and forth at
right angles along rods of length $L$ (at rest in the S' frame)
 \BE \label{deltaT} \Delta t(\theta)=
(2L/\bar{c}_\gamma(\theta))- (2L/\bar{c}_\gamma(\pi/2+\theta))\sim
(2L/c) (\Delta\bar{c}_\theta /c)\EE This gives back the
phenomenologically successful Eqs.(\ref{fringe}) and (\ref{vobs}).
All together, we have found a consistent description of the data
where symmetry arguments, on the one hand, motivate and, on the
other hand, find justification in underlying dynamical mechanisms.

\section{Conclusions and outlook}

\begin{figure}
\begin{center}
\epsfig{figure=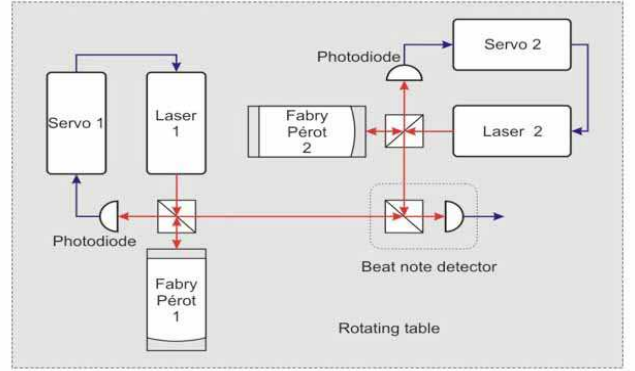,width=8.5 truecm,angle=0}
\end{center}
\caption{ {\it The scheme of a modern ether-drift experiment. The
light frequencies are first stabilized by coupling the lasers to
Fabry-Perot optical resonators. The frequencies $\nu_1$ and $\nu_2$
of the signals from the resonators are then compared in the beat
note detector which provides the frequency shift $\Delta \nu=\nu_1
-\nu_2$. In present experiments a very high vacuum is maintained
within the resonators. } } \label{apparatus}
\end{figure}

This overall level of consistency requires a check with a new
generation of precise laser interferometers in order to reproduce
the experimental conditions of the old experiments with today's much
greater accuracy. The essential ingredient is that the optical
resonators that nowadays are coupled to the lasers should be filled
by gaseous media, see Fig.\ref{apparatus}.
Such a type of `non-vacuum' experiments would be along the lines of
ref.\cite{holger} where just the use of optical cavities filled with
different materials was considered as a useful complementary tool to
study deviations from exact Lorentz invariance. The only delicate
aspect concerns the high relative stability in temperature and
pressure of the two cavities which is required to prevent possible
spurious sources of the frequency shifts. However, with present
technology and technical skill \footnote{For instance, an important
element to increase the overall stability and minimize systematic
effects may consist in obtaining the two optical resonators from the
same block of material as with the crossed optical cavity of
ref.\cite{crossed}.}, this should not represent a too serious
problem.
\begin{figure}
\begin{center}
\includegraphics [scale=0.2] 
{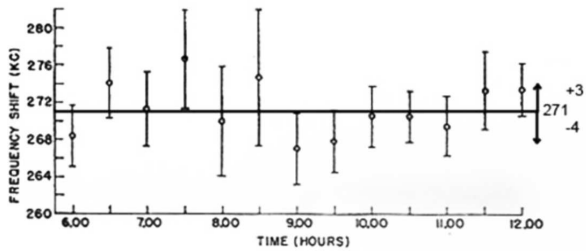} \caption{\it The frequency shifts of
ref.\cite{jaseja}. The double arrow indicates the overall variation,
with respect to the constant value, expected in the same model
Eqs.(\ref{twoway0}) and (\ref{anigas}) used for the classical
experiments. To this end, we have assumed the range $v \sim
320^{+45}_{-60}$ km/s, as for the average CBR Earth's motion at the
latitude of Boston. In this way, once the shift for the average
velocity 320 km/s is hidden in the much larger spurious frequency
shift, the two relative variations of $+3$ kHz and $-4$ kHz would
correspond respectively to $v=$365 and $v=$260 km/s. }\label{mit}
\end{center}
\end{figure}

In units of their natural frequency $\nu_0$, we then predict a
frequency shift between the two resonators \BE \label{anigas}
(\Delta\nu/\nu_0)_{\rm gas} =(\Delta\bar{c}_\theta/c)_{\rm gas}\sim
({\cal N}_{\rm gas} -1)~(v^2/c^2) \EE which should be larger by
orders of magnitude than the corresponding effect with vacuum
resonators \cite{brillet}$-$\cite{schillernew}. This substantial
enhancement is confirmed by the only modern experiment that has been
performed in similar conditions: the 1963 MIT experiment by Jaseja
et. al \cite{jaseja}. They were looking at the frequency shift of
two orthogonal He-Ne lasers placed on a rotating platform. For a
proper comparison, one has to subtract preliminarily a large
systematic effect of about 270 kHz interpreted as being due to
magnetostriction. As suggested by the same authors, this spurious
effect, that was only affecting the normalization of the
experimental $\Delta \nu$, can be subtracted by looking at the
variations of the data. In this case, for a laser frequency
$\nu_0\sim 2.6 \cdot 10^{14}$ Hz, the residual variations of a few
kHz, see Fig.\ref{mit}, are roughly consistent with the refractive
index ${\cal N}_{\rm He-Ne}\sim 1.00004$ and the typical variations
of the Earth's velocities in Table 1.

To conclude, suppose some future experiment would confirm the
unambiguous detection in gaseous systems of a universal signal as
given by Eq.(\ref{anigas}). This could have other non-trivial
implications. In fact, it would mean that all physical systems on
the moving Earth are exposed to a tiny energy flow, an effect which,
in principle, could induce forms of spontaneous self-organization in
matter \cite{prigogine,soc2}. In slightly different terms, the
existence of such a flow introduces a weak, residual form of `noise'
which is intrinsic to natural phenomena (`objective noise'
\cite{grigolini}). This could be crucial because it has becoming
more and more evident that many classical and quantum systems can
increase their efficiency thanks to the presence of noise (e.g.
photosynthesis in sulphur bacteria \cite{caruso1}, quantum transport
\cite{caruso2}, protein crystallization \cite{frenkel}, noise
enhanced stability \cite{spagnolo} or stochastic resonance
\cite{gamma1}). In this sense, a fundamental signal with genuine
characters of turbulence or intermittency could be thought as the
microscopic origin of macroscopic aspects such as self-organized
criticality, large-scale fluctuations, fat-tailed probability
density functions among many others, which characterize the behavior
of many complex systems, see e.g.
\cite{sreenivasan,beck,beck-cohen,tsallis}.

\end{document}